\documentclass[aps,pra,twocolumn,groupedaddress,showpacs,superscriptaddress,amssymb,amsmath]
{revtex4-2}
\usepackage{physics}
\usepackage{pgfplots}
\pgfplotsset{compat=1.13}
\usepackage[utf8]{inputenc}
\usepackage{graphicx}
\usepackage{tabularx}
\usepackage{xcolor}
\usepackage{amsmath}
\usepackage{comment}
\usepackage{dcolumn}
\usepackage{hyperref}
\hypersetup{
    colorlinks=true,
    linkcolor=blue,
	citecolor=cyan
}
\usepackage{bm}
\usepackage{epsf}
\usepackage{braket}
\usepackage{tensor}
\usepackage{soul}
\usepackage{float}
\usepackage{subfig}
\usepackage{mathrsfs}
\usepackage{mathtools}
\usepackage{multirow}
\usepackage{caption}
\usepackage{amsfonts}

\newcommand{\be}{\begin{equation}}
\newcommand{\ee}{\end{equation}}
\newcommand{\bea}{\begin{eqnarray}}
\newcommand{\eea}{\end{eqnarray}}

\makeatletter
\newsavebox{\@brx}
\newcommand{\llangle}[1][]{\savebox{\@brx}{\(\m@th{#1\langle}\)}%
  \mathopen{\copy\@brx\kern-0.5\wd\@brx\usebox{\@brx}}}
\newcommand{\rrangle}[1][]{\savebox{\@brx}{\(\m@th{#1\rangle}\)}%
  \mathclose{\copy\@brx\kern-0.5\wd\@brx\usebox{\@brx}}}
\makeatother

\begin{document}
\title{Bootstrapping non-Hermitian Quantum System}
\author{Sakil Khan}
\email{sakil.khan@students.iiserpune.ac.in}
\affiliation{Department of Physics,
		Indian Institute of Science Education and Research, Pune 411008, India}
\author{Harsh Rathod}
\email{rathod.harsh@students.iiserpune.ac.in}
\affiliation{Department of Physics,
		Indian Institute of Science Education and Research, Pune 411008, India}
\date{\today}
\begin{abstract}
Recently, the ``Bootstrap" technique was applied in Quantum Mechanics to solve the eigenspectra of Hermitian Hamiltonians and extended to non-Hermitian PT-symmetric systems. However, its application has been limited to real spectra. In this work, we establish bootstrap conditions for the non-Hermitian system and generate eigenspectra for a generic complex polynomial potential, which includes PT-symmetric Hamiltonians as a special case. Additionally, we demonstrate the method's ability to obtain eigenspectra under various boundary conditions imposed on the eigenfunction, including the notable application of capturing the PT-symmetric phase transition.
\end{abstract}
  \maketitle  

\section{Introduction}
In the ``Bootstrap" technique, fundamental constraints such as unitarity, causality, and positivity are used to make rigorous numerical or analytical statements about observables in quantum theories. This approach has been widely applied in  Quantum Field Theory (QFT) and Conformal Field Theory (CFT) \cite{1974JETP...39...10P,El-Showk:2012cjh,rattazzi2008bounding,poland2019conformal}. More recently, it has been used in Quantum Mechanics  \cite{hartnol}.

In the context of quantum mechanical systems, this technique has been used to solve the eigenspectra of Hermitian Hamiltonians \cite{PhysRevD.105.085017,aikawa2022bootstrap,berenstein2024numerical,PhysRevD.106.045029,PhysRevD.106.116008,Bai:2022yfv,Berenstein:2021dyf,Bhattacharya:2021btd,Du:2021hfw,kazakov2022analytic,Lawrence:2021msm,Nakayama:2022ahr,Tchoumakov:2021mnh,berenstein2022bootstrapping,nancarrow2023bootstrapping,PhysRevD.109.025013,du2022bootstrapping,PhysRevD.107.094511,PhysRevD.109.126002,john2023anharmonic,fan2023non,hu2022different,han2020quantum,fan2023non} and has been further extended to non-Hermitian PT-symmetric systems in subsequent studies \cite{KHAN2022137445,PhysRevD.106.125021}. The key principle underlying the bootstrap method is the positivity of the state's norm or the unitarity. However, the development of bootstrapping quantum systems has been limited to the real spectra. One genuine question is how this technique can be applied to complex spectra. 
Complex spectra naturally occur in open quantum systems \cite{breuer,Carmichael,lidar2019lecture,davies}, where a system interacts with its environment. Studying these systems is crucial because they more accurately reflect realistic conditions, and their spectra can reveal important physical characteristics, such as how the system thermalizes. One of the primary challenges we face is identifying appropriate constraints for bootstrapping such systems.


\textbf{Complex spectra:}
As we mentioned above, Positivity or Unitarity can serve as constraints for Hermitian systems.
For non-Hermitian systems, the positivity constraint still applies, as the norm (with respect to complex conjugation) of any state must be positive. This is a fundamental property of Hilbert space. In the Hermitian case, along with the positivity constraint, two equality-type constraints are also required: $ \langle{ E_{n}}|[H,O]|E_{n} \rangle =0$ and $ \langle{ E_{n}}|OH|E_{n} \rangle =E_{n}\langle{ E_{n}}|O|E_{n} \rangle $. However, these constraints do not apply to non-Hermitian systems \cite{KHAN2022137445,PhysRevD.106.125021}. In this work, we identify appropriate bootstrap conditions for a generic non-Hermitian system and use this technique to generate the eigenspectra for any complex polynomial potential. 
Additionally, we illustrate the effectiveness of our bootstrap technique in generating different eigenvalues under various boundary conditions imposed on the eigenfunction  \cite{Bender:1992bk,Bender_2007} using a simple example. Below, we outline one interesting application of the developed method.

\textbf{PT-symmetric Phase Transition:}
If a non-Hermitian Hamiltonian is invariant under the PT operator and its eigenvector respects PT symmetry, the corresponding eigenvalue will be real \cite{benderprl,ptbender,NOBLE2017304,JOUR,-x4,V,swanson}. In contrast, if the eigenstate does not respect PT symmetry, the eigenvalues become complex, and we say that PT symmetry is broken \cite{Bender_2007}. This transition from real to complex eigenvalues is known as the PT-symmetric phase transition. Since our developed method can handle both real and complex spectra, we demonstrate that it is ideally suited to capture this phase transition. 

We organize the paper as follows: In Section \ref{sec2}, we develop the bootstrap technique for non-Hermitian systems and present a systematic framework for generating the spectra of any complex polynomial potential. We apply the bootstrap method to PT-symmetric systems, a subclass of non-Hermitian systems in Section \ref{sec3}. In Section \ref{sec4}, we generate complex spectra for specific non-Hermitian potentials and also demonstrate the effectiveness of the bootstrap technique in addressing problems with various boundary conditions.
We show one important application of the method, illustrating the PT-symmetric phase transition in Section \ref{sec6}.
  Finally, we summarize our findings in Section \ref{sec7}. Additional details, including the full derivation of the bootstrap conditions, are provided in the appendix.


\section{Bootstrapping non-Hermitian potential}
In this section, we will begin by developing the bootstrap technique for non-Hermitian systems. Next, we will present a general framework for generating the eigenspectra of any complex polynomial potential. We will then implement the bootstrap technique to a PT-symmetric system using a specific example. Finally, we will generate complex spectra to illustrate the applicability and efficiency of the developed technique in a broader non-Hermitian context, using a few examples.
\subsection{Methodology}\label{sec2}
In a Hermitian system, the left and right eigenvectors are identical, making it straightforward to define orthonormality conditions. However, in a non-Hermitian system, the left and right eigenvectors are generally distinct. Let us assume $|R_{n}\rangle$ and $\langle L_{n}|$ represent the 
nth right and left eigenstates, respectively i.e.
\begin{align}
	&H|R_{n} \rangle=(E^{n}_{\mathcal{R}}+i E^{n}_{\mathcal{I}})|R_{n} \rangle\nonumber\\
 &\langle L_{n}|H=(E^{n}_{\mathcal{R}}+i E^{n}_{\mathcal{I}})\langle L_{n}|
\end{align}
Note that the spectra of the Hamiltonian can generally be complex, with $E^{n}_{\mathcal{R}}$ and $E^{n}_{\mathcal{I}}$ representing the real and imaginary parts of the 
nth eigenvalue, respectively. If we define $\langle R_{n}|$ as the complex conjugate of $|R_{n}\rangle$, then, unlike in Hermitian systems, in non-Hermitian systems, $\langle R_{m}|R_{n}\rangle \neq \delta_{mn}$, indicating that they are not orthonormal. Consequently, the usual bootstrap conditions do not apply to these systems. However, bootstrapping can still be performed using the following constraints:
\begin{align}\label{bcn}
	&  \langle{ R_{n}}|O^{\dagger}O|R_{n} \rangle\geq 0
 \nonumber\\
	&    \langle{ R_{n}}|(OH-H^{\dagger}O)|R_{n} \rangle =2 iE^{n}_{\mathcal{I}}\langle{ R_{n}}|O|R_{n} \rangle\nonumber\\
	&  \langle{ R_{n}}|(OH)|R_{n} \rangle =(E^{n}_{\mathcal{R}}+i E^{n}_{\mathcal{I}})\langle{ R_{n}}|O|R_{n} \rangle \;.
\end{align}
A detailed derivation of the above bootstrap condition is provided in the Appendix \ref{app1}. Note that for the Hermitian system, where $H^\dagger=H$ and the eigenspectra are real, the above bootstrap conditions reduce to the standard bootstrap conditions \cite{hartnol}. Below, we outline a general framework for obtaining the eigenspectra of a generic complex polynomial potential.
\subsection*{Complex Polynomial Potential}
Complex polynomial potentials naturally arise in the context of open quantum systems. The Hamiltonian for a complex polynomial potential can be expressed as:
\begin{equation}\label{eqp5.0}
	H=p^2+V_{1}(x)+ i\;
 V_{2}(x)\;,
\end{equation}
where $V_{1}(x)$, $V_{2}(x)$ are polynomial function of $x$ and are also  Hermitian operators. Since $V_{1}(x)$, $V_{2}(x)$ are polynomial potential, they can be represented as $V_{1}(x)=\sum^{d_{1}}_{n=1} a_{n}x^n$ and $V_{2}(x)=\sum^{d_{2}}_{m=1} b_{m}x^m$.
If we choose $O=\sum_{t}\alpha_{t} x^t$,  the positivity constraints imply that the matrix 
$M$ must be positive semi-definite i.e. $M\geq 0$ with $M_{ij}=\langle{ R_{n}}|x^{i+j}|R_{n} \rangle$.

Using Eq.\eqref{bcn}, we can derive the following recursion relation between the moments of 
 $x$ \big($ \langle x^{t}\rangle= \langle{ R_{n}}|x^t|R_{n} \rangle$\big):
 \small
\begin{align}\label{recn}
	&2 t E_{\mathcal{R}}\langle x^{t-1}\rangle-\sum^{d_{1}}_{n=1}(2t+n) a_{n}\langle x^{t+n-1}\rangle+t(t-1)(t-2)\frac{\langle x^{t-3}\rangle}{2}\nonumber\\
 &+\sum^{d_{2}}_{m=1}\frac{ 2b_{m}E_{\mathcal{I}}}{(t+1)}\langle x^{t+m+1}\rangle-\frac{ 2E^{2}_{\mathcal{I}}}{(t+1)}\langle x^{t+1}\rangle +\sum^{d_{2}}_{m=1}b_{m}\Big(   \frac{2E_{\mathcal{I}}}{(t+m+1)}\nonumber\\
 &    \langle x^{t+m+1}\rangle -\sum^{d_{2}}_{k=1}\frac{2 b_{k}}{(t+m+1)} \langle x^{t+m+k+1}\rangle      \Big)=0\;.
\end{align}
\normalsize
The above recursion relation is crucial as it expresses higher moments of 
$x$ in terms of lower moments. 
It's important to note that when the number of unknown parameters—such as the real and imaginary parts of the eigenvalue and the independent moments of 
$x$—exceeds three, finding these parameters becomes computationally challenging. By using semidefinite programming (SDP) \cite{PhysRevE.107.L053301,simmons2015semidefinite}, we significantly increase the computational efficiency of the problem. We illustrate this method using several examples. Below, we will first bootstrap a PT-symmetric non-Hermitian Hamiltonian.

\subsection{Example of PT-symmetric non-Hermitian potential and implementation of bootstrap}\label{sec3}
\begin{figure}
	\centering
\includegraphics[scale=.059]{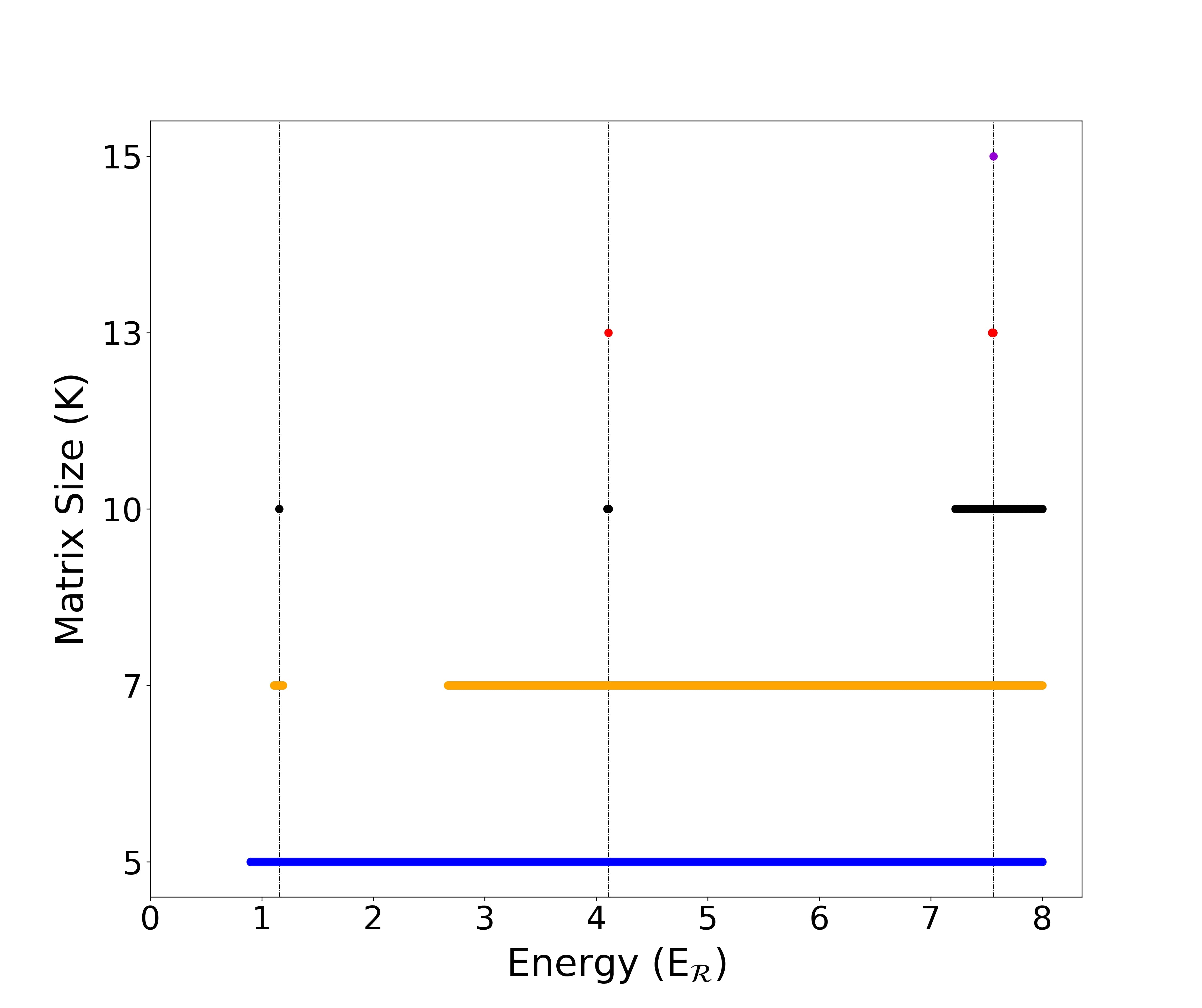}
	\caption{(Color online:) We have plotted the first three energy eigenvalues ($E_{\mathcal{R}}$) for the 
PT-symmetric $ix^3$  potential, using various bootstrap matrix sizes 
$K$. Note that, the eigenvalues are real for this potential.
The plot shows that the eigenvalues converge quickly as the matrix size 
$K$ increases. However,  as we move to higher states, the bootstrap matrix sizes also need to be increased to maintain the same level of accuracy in determining the energy values.}
	\label{fig:1}
\end{figure}

PT-symmetric non-Hermitian Hamiltonians form a unique subclass of non-Hermitian systems that can exhibit real spectra. In a previous work \cite{KHAN2022137445}, we developed a bootstrapping method, but it was limited to generating eigenspectra for only specific PT-symmetric potentials. For a more detailed discussion, see the footnote \footnote{The bootstrap method we developed in \cite{KHAN2022137445} for PT-symmetric systems depends on the $V$ operator, defined by the following conditions: 
$ VH V^{-1}=H^\dagger,
  V^\dagger=V$,
and $V \text{ is a positive operator}  $. However, identifying the $V$ operator for a given Hamiltonian is not always feasible. In fact, for the $ix^3$ potential, it can only be determined perturbatively. 
}. However, our current work not only addresses this limitation but extends the approach to any non-Hermitian system. To illustrate this, we will compute the eigenvalues of the PT-symmetric $ix^3$ potential—a special class of non-Hermitian potential—that could not be solved by the method in \cite{KHAN2022137445} \footnote{Note that the 
$ix^3$ potential was solved using a different method (null bootstrap) in \cite{PhysRevD.106.125021}. However, our current method is a direct generalization of the original approach \cite{hartnol}, making it more natural and intuitive.}.
The Hamiltonian is given by
\begin{equation}\label{ix3}
	H=p^2+ix^3\;.
\end{equation}
The Hamiltonian above is PT-symmetric and has real spectra. By applying the recursion relation in Eq.\eqref{recn}, we can express higher moments of 
$x$ in terms of lower moments, thus reducing the number of unknown parameters. The explicit recursion relation is given by
\begin{align}
	&4 \langle x^{t+7}\rangle-4t(t+4)E_{\mathcal{R}}\langle x^{t-1}\rangle\nonumber\\
 &-t(t-1)(t-2)(t+4)\langle x^{t-3}\rangle=0\;.
\end{align}
For this potential, the search space is seven-dimensional \footnote{For this potential, the unknowns are real part of the eigenvalue and six moments of $x$}. Since the recursion relation separates odd and even moments of 
$x$, focusing only on the even moments in the bootstrap matrix reduces the number of unknown parameters to four. Given that the space of unknown parameters is four, we do not scan all parameters but instead use semidefinite programming (SDP) to only search over the energy eigenvalue. By employing the SDP method, we numerically determine the eigenvalues of 
$H$ defined in Eq.\eqref{ix3}. The bootstrap results are presented in Table \ref{tab:t1}
, where 
$K$ denotes the size of the matrix 
$M$ which is used for the positivity constraint.
We have plotted the possible energy values for the ground state, first excited state, and second excited state in Fig.\ref{fig:1}. 
The figure shows that the allowed energy eigenvalues converge quickly with increasing bootstrap matrix size $K$. However, note that as we move to higher states, the bootstrap matrix sizes also need to be increased to maintain the same level of accuracy in determining the energy values.

    \begin{table}[h!]
    \centering
\begin{tabular}{ |p{2cm}||p{2cm}|p{2cm}|p{2cm}|  }
 \hline
 \multicolumn{4}{|c|}{Spectrum for $p^{2}+ix^{3}$} \\
 \hline
 K& 7& 10 & $\textit{Python}$ \\
 \hline
   
  Ground State (Energy)&  [1.110, 1.190] & 1.156 &1.156\\
 & & &  \\
 \hline
 K& 10&13 &$\textit{Python}$\\
 \hline
  First Excited State  & [4.100, 4.110]  &  4.110&4.110\\
 & & & \\
 \hline
 K& 13&15 &$\textit{Python}$\\
 \hline
  First Excited State  & [7.550, 7.560  ]&  7.562&7.562\\
 & & & \\
 \hline
\end{tabular}
\captionof{table}{This table displays the range of energy eigenvalues obtained through bootstrapping for the ground state, first excited state, and second excited state of the PT-symmetric $ix^3$ potential (real eigenspectra), across various matrix sizes 
$K$. We used square brackets [ ] to indicate the allowed range of energy values.
This table shows a nice convergence of the eigenvalues with $K$. Eigenvalues, obtained by numerically solving (using RK method) the Schrödinger equation, are also provided.}
\label{tab:t1}
\end{table}
\subsection{Bootstrapping complex spectra}\label{sec4}
We bootstrapped the PT-symmetric system in the previous subsection. In this section, we will generate complex spectra to demonstrate the applicability and efficiency of the developed technique in a general non-Hermitian regime. Furthermore, we illustrate the effectiveness of our bootstrap technique in generating different eigenvalues under various boundary conditions imposed on the eigenfunction using a simple example. 
We will first explore the following two examples.
\subsubsection*{Example:1}
\begin{figure}
	\subfloat[]{\includegraphics[scale=.25]{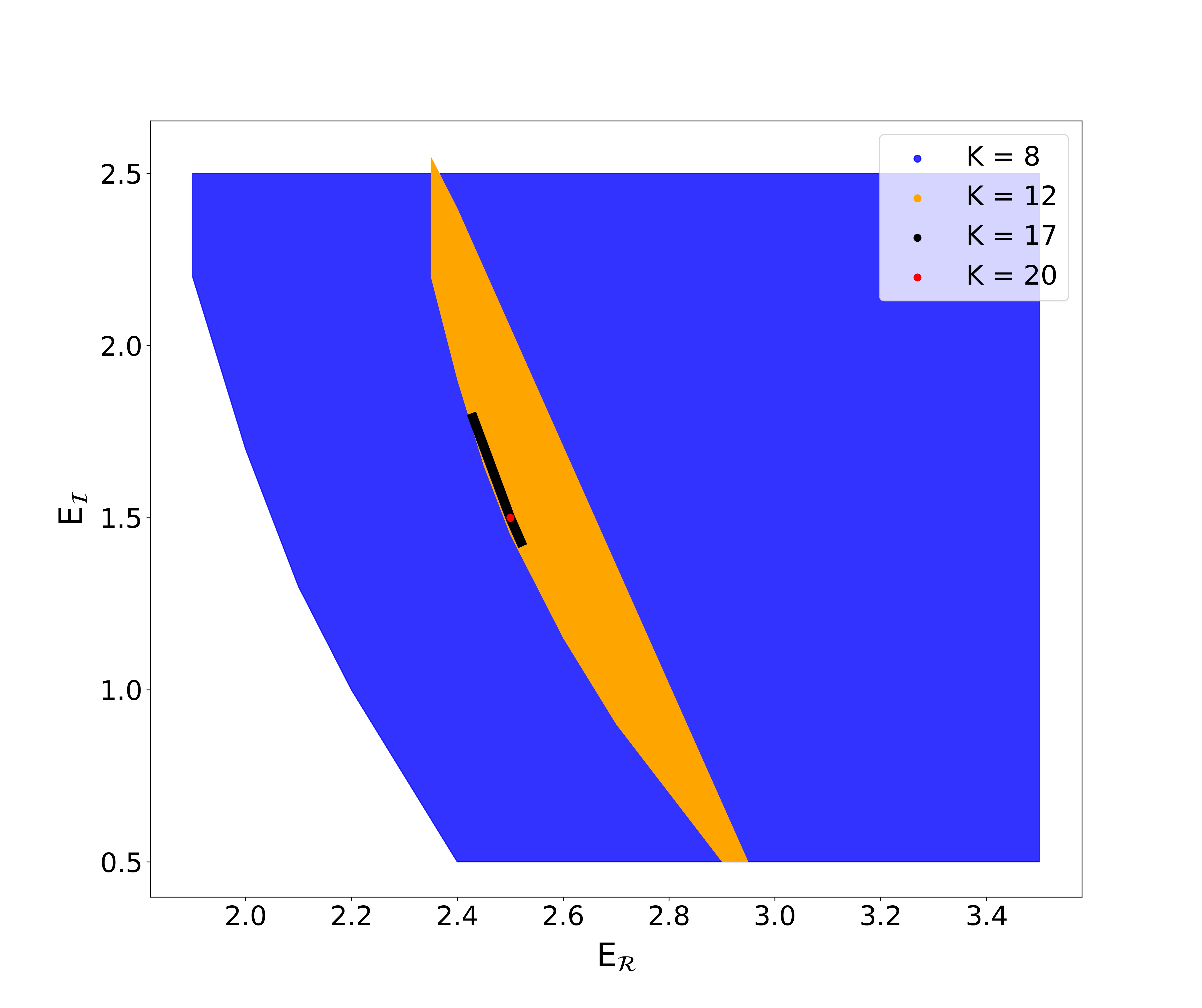}}\\
 
	\subfloat[]{\includegraphics[scale=.23]{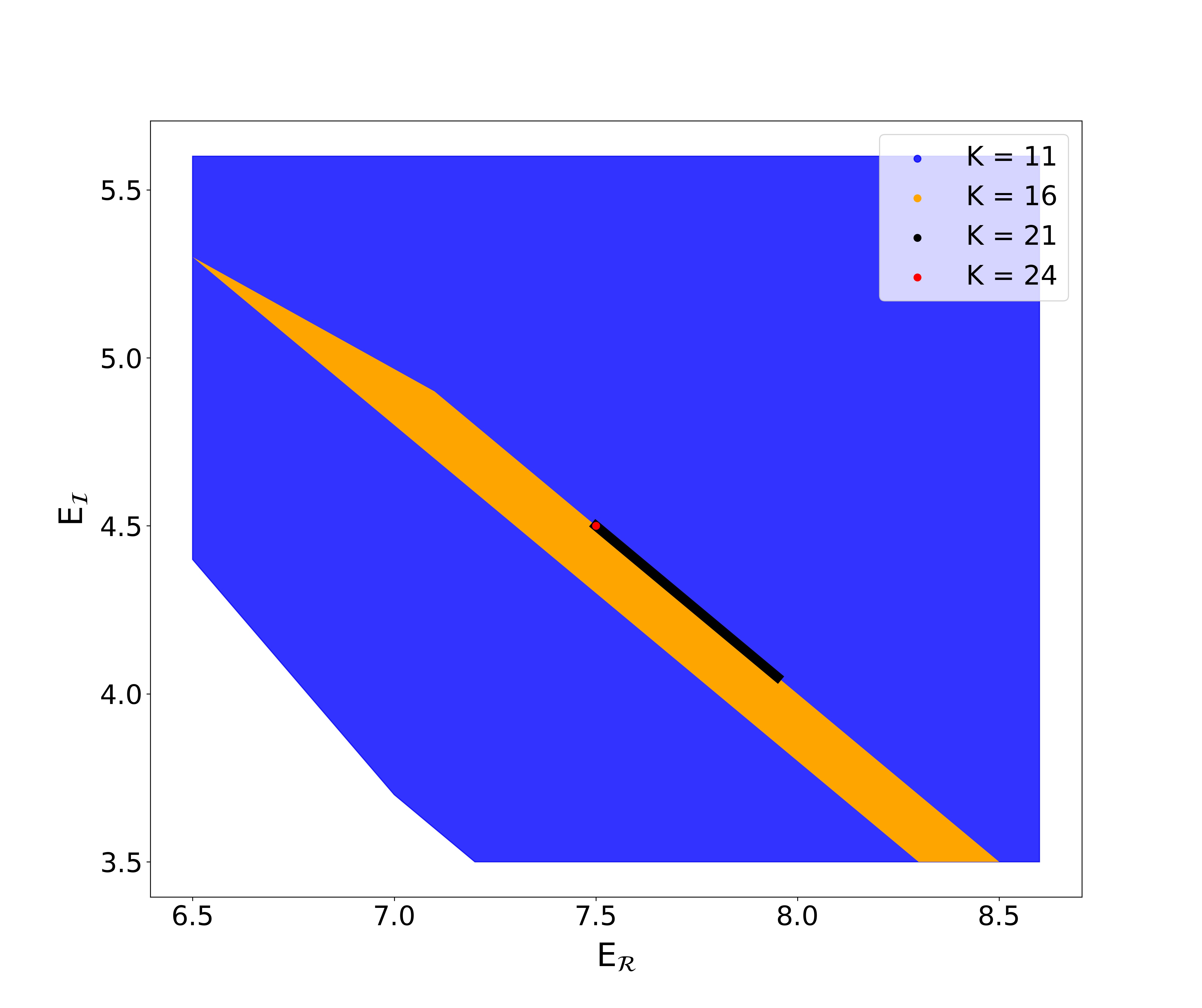}}
    \caption{(Color online:) We have plotted the first and second energy eigenvalues for the complexified harmonic potential (which is not PT-symmetric), using various bootstrap matrix sizes 
$K$ [(a)-(b)]. Note that, the eigenvalues are complex for this potential.
The plot shows that both the real and imaginary parts of the eigenvalues converge quickly as the matrix size 
$K$ increases. }
	\label{fig:2}
\end{figure}

First, we consider a simple example: the complexified version of the simple harmonic oscillator. The Hamiltonian is given by \cite{Jannussis1986HarmonicOW}
\begin{equation}\label{eqp5.0}
	H=p^2+(\omega_{1}+i\omega_{2})^2 \frac{x^2}{4}\;,
\end{equation}
where $\omega_{1}$, $\omega_{2}$
  are real parameters. This Hamiltonian is clearly not PT-symmetric and has a complex spectrum. It can even be solved analytically, resulting in the energy eigenvalues $E=(2n+1)(\omega_{1}+i\omega_{2})/2$ \cite{Jannussis1986HarmonicOW}.
  In Table  \ref{tab:t2}, we present the bootstrap results and compare them with the exact solutions. The possible energy eigenvalues for the first two states are plotted in Fig.\ref{fig:2}. Note that, the eigenvalues are complex for this potential.
The plot shows that both the real and imaginary parts of the eigenvalues converge quickly as the matrix size 
$K$ increases.
 \begin{table}[h!]
    \centering
\begin{tabular}{ |p{2cm}||p{2cm}|p{2cm}|p{2cm}|  }
 \hline
 \multicolumn{4}{|c|}{Spectrum for $p^{2}+(5 +3i)\frac{x^2}{4}$} \\
 \hline
 K& 12& 20 & Exact Value \\
 \hline
   
  1st State&$E_{\mathcal{R}}:[2.3,2.9]$
  $E_{\mathcal{I}}:[0.5, 2.5] $ & 2.50+1.50$i$&2.5+1.5$i$\\
 & & &  \\
 \hline
 K& 16&24 &Exact Value\\
 \hline
 
 2nd State &$E_{\mathcal{R}}:[6.5,8.5]$
  $E_{\mathcal{I}}:[3.5,5.3] $   & 7.50+4.50$i$&7.5+4.5$i$\\
 & & & \\
 \hline
 
\end{tabular}
\captionof{table}{This table shows the range of energy eigenvalues obtained through bootstrapping for the first two states of the complexified harmonic potential (which is not PT-symmetric), across various matrix sizes 
$K$. We used square brackets [ ] to indicate the allowed range of energy values.
For this potential the eigenvalues are complex.
 Exact values, calculated by analytically solving the Schrödinger equation, are also provided.}
\label{tab:t2}
\end{table}

\subsubsection*{Example:2}
\begin{figure}
	\subfloat[]{\includegraphics[scale=.23]{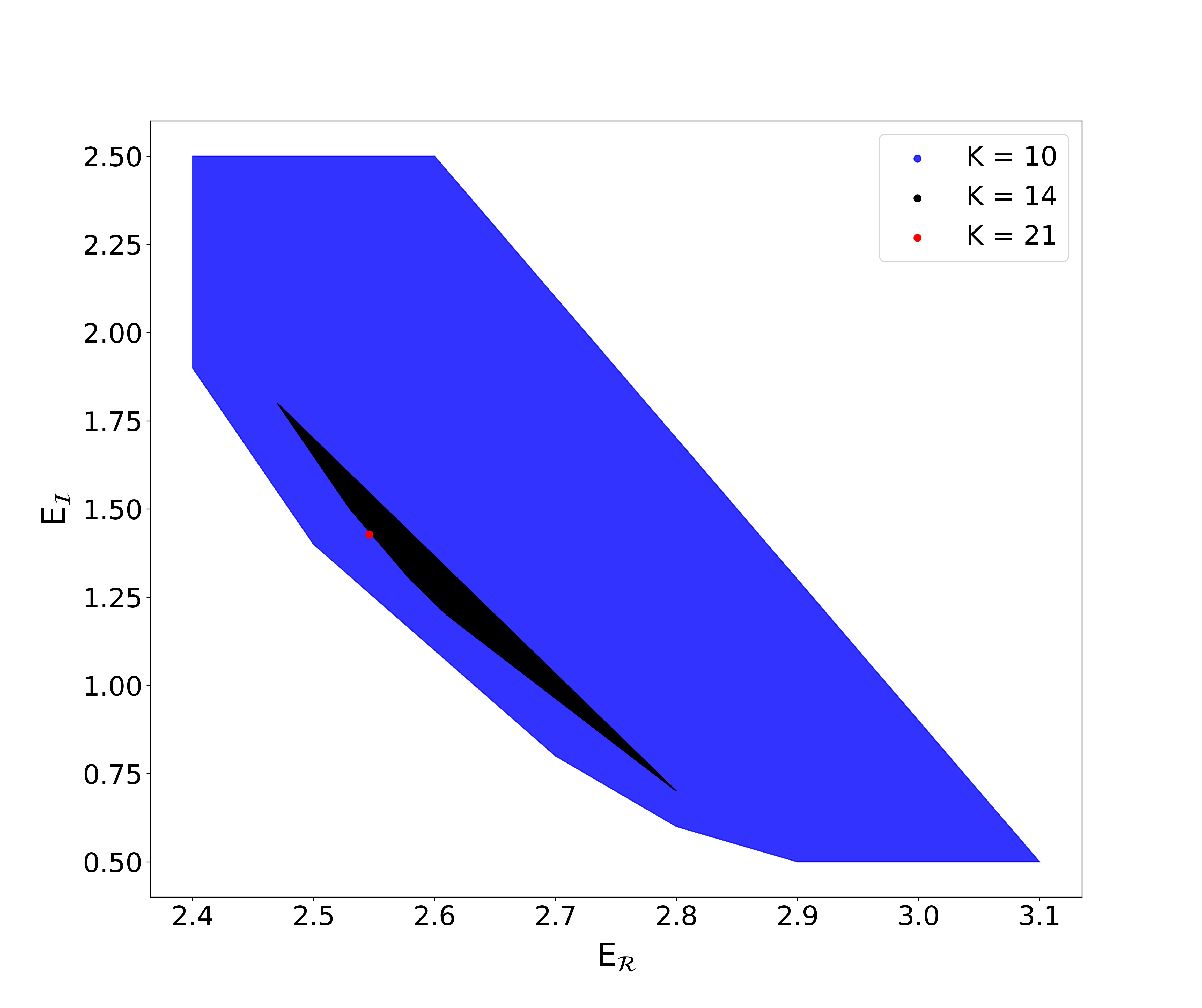}}\\
 
	\subfloat[]{\includegraphics[scale=.23]{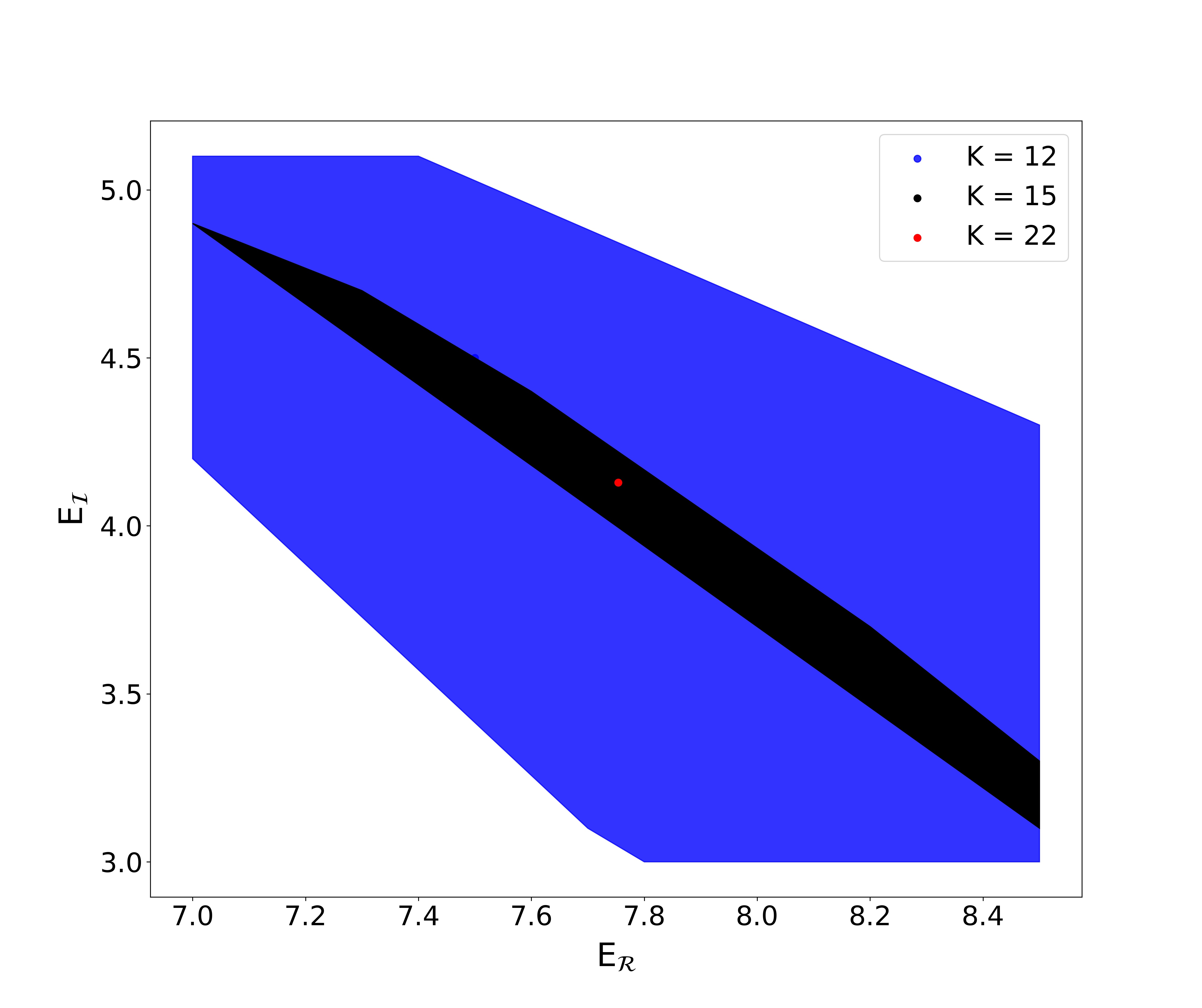}}
    \caption{(Color online:) We have plotted the first and second energy eigenvalues for the complexified anharmonic potential, utilizing different bootstrap matrix sizes $K$ [(a)-(b)]. The eigenvalues are again complex for this potential. Note that, this potential is not solvable analytically. The plot shows that both the real and imaginary parts of the eigenvalues converge very fast as the matrix size 
$K$ increases.}
	\label{fig:3}
\end{figure}

Next, we consider the complexified version of the anharmonic oscillator, which cannot be solved analytically. The Hamiltonian is given by
\begin{equation}\label{eqp5.0}
	H=p^2+(\omega_{1}+i\omega_{2})^2 \frac{x^2}{4}+g x^4\;,
\end{equation}
By using the recursion relation outlined in Eq.\eqref{recn}, we can determine that the total number of unknown parameters for this potential is six. We compute the first two eigenvalues for this potential and compare them with results obtained using another method (Python). The outcomes are presented in Table \ref{tab:t3}
. The possible energy values for the first two states are plotted in Fig.\ref{fig:3}. The eigenvalues are again complex for this potential. As mentioned above, this potential is not solvable analytically. The plot shows that both the real and imaginary parts of the eigenvalues converge quickly as the matrix size 
$K$ increases. Below, we compare two distinct sets of eigenvalues that emerge from the two different boundary conditions imposed on the wavefunction, illustrated through a simple example.
 \begin{table}[h!]
    \centering
\begin{tabular}{ |p{2cm}||p{2cm}|p{2cm}|p{2cm}|  }
 \hline
 \multicolumn{4}{|c|}{Spectrum for $p^{2}+(5 +3i)\frac{x^2}{4}+x^4$} \\
 \hline
 K& 14& 21 & $\textit{Python}$ \\
 \hline
   
  First State& $E_{\mathcal{R}}:[2.5,2.8]$
  $E_{\mathcal{I}}:[0.7, 1.8] $   & 2.55+1.43$i$&2.55+1.43$i$\\
 & & &  \\
 \hline
 K& 15&22 &$\textit{Python}$\\
 \hline
 
 Second State & $E_{\mathcal{R}}:[7.0,8.5]$
  $E_{\mathcal{I}}:[3.1, 4.9] $ & 7.75+4.14$i$&7.75+4.14$i$\\
 & & & \\
 \hline
 
 
\end{tabular}
\captionof{table}{The table provides the range of energy eigenvalues for the first two states of the complexified anharmonic potential (eigenvalues are again complex for this potential), as determined by bootstrapping across different matrix sizes 
$K$. We used square brackets [ ] to indicate the allowed range of energy values.
It also includes eigenvalues obtained from numerically solving (Python) the Schrödinger equation. }
\label{tab:t3}
\end{table}
\subsubsection*{Complex boundary condition:}\label{sec5}
The spectrum of a Hamiltonian is heavily dependent on the boundary conditions imposed on the eigenfunction. The standard boundary condition requires the eigenfunction to vanish at $\pm \infty$. However, by analytically continuing the Schrödinger equation into the complex plane, we can obtain different sets of eigenvalues based on various boundary conditions imposed on the eigenfunction. Interestingly, a system that exhibits real spectra under standard conditions may produce complex spectra under alternative conditions. For example, consider the following Hamiltonian:
\begin{equation}\label{}
	H=p^2+ x^6-15 x\;.
\end{equation}
The spectra of the above Hamiltonian are real when the standard boundary condition is applied to the eigenfunction. However, by imposing the boundary condition along the imaginary axis, meaning the eigenfunction vanishes at $\pm i \infty$, we obtain complex spectra. To illustrate this, let's first substitute  $x=i r$, which transforms the Hamiltonian into:
\begin{equation}\label{}
	H=-(p^{2}_{r}+r^6 +15 ir)\;,
\end{equation}
where $[r,p_{r}]=i$. By applying the bootstrap technique, we find that the above Hamiltonian exhibits complex spectra. In Table \ref{tab:t4}, we compare the eigenvalues obtained under the different boundary conditions. The possible energy values for the first two states are plotted in Fig.\ref{fig:4}.

\begin{table}[h!]
    \centering
\begin{tabular}{ |p{1.5cm}||p{1.5cm}|p{1.5cm}|p{1.5cm}||p{1.5cm}|  }
 \hline
 \multicolumn{5}{|c|}{Comparison of two different boundary condition } \\
 \hline
State & Standard boundary condition & K & Deformed boundary condition & K\\
 \hline
    First &   -9.72 & 16&-8.88 $\pm10.84i$&17\\
 & & & & \\
 \hline
  Second &   -0.63 & 19&-15.50 $\pm3.30i$&22\\ & & & &\\
 \hline
\end{tabular}
\captionof{table}{This table compares the eigenvalues obtained through bootstrapping under two different boundary conditions. Real eigenvalues are observed under the standard boundary condition, whereas complex eigenvalues appear under the deformed boundary condition.}
\label{tab:t4}
\end{table}

So far, we have developed the bootstrap technique for non-Hermitian systems and demonstrated its effectiveness in generating both real and complex spectra through various examples. Next, we will present one important application of our method.

\section{PT-Symmetric Phase Transition}\label{sec6}
 The PT-symmetric non-Hermitian system may exhibit real spectra. Consider that 
 $|\psi\rangle$  is an eigenstate of both the Hamiltonian 
$H$ and the  $PT$ operator. Since the $PT$ operator commutes with the Hamiltonian, we can write the following 
\begin{align}
	&[PT, H]  |\psi\rangle=0\nonumber\\
&\text{or,} \;\;\lambda( E -E^* )=0\;,
\end{align}
where $E$ and  $\lambda$  are the eigenvalues of $H$ and the $PT$ operator, respectively. It can be shown that the eigenvalue of the PT operator ($\lambda$) is a pure phase and therefore non-zero \cite{Bender_2007}, which implies that 
$E$ must be real, i.e.,  $E=E^*$.
It’s important to note that, in proving 
$E$ is real, we assumed 
$|\psi\rangle$  is an eigenstate of both 
$H$ and the $PT$ operator, given that $PT$ commutes with 
$H$. However, this assumption is not always valid because $PT$ is an anti-linear operator. If the eigenstate of a PT-symmetric Hamiltonian does not respect PT symmetry, the corresponding eigenvalue becomes complex, a condition referred to as PT symmetry breaking. In this section, we will illustrate this concept with a simple example.

We consider a particle in a quartic potential and introduce a linear non-Hermitian term to make the system non-Hermitian and PT-symmetric. The Hamiltonian is given by \cite{Handy}:
\begin{equation}\label{}
	H=p^2+ x^4 + i a x\;.
\end{equation}
Although this Hamiltonian is PT-symmetric, the extent to which the eigenstates respect PT symmetry depends on the value of the parameter 
$a$ \cite{Handy}. We examined various values of 
$a$ and analyzed the nature of the ground state using the bootstrap technique. Detailed results are provided in Table \ref{tab:t5}.  We have plotted the ground state energy eigenvalues for the PT-symmetric $x^4+iax$ potential with 
$a=1$ and 
$a=3$, using various bootstrap matrix sizes 
$K$ in Fig.\ref{fig:5}.  For these particular values of $a$, the eigenfunction of the system respects the PT-symmetry, ensuring the real spectra.  In Fig.\ref{fig:6}, we have plotted the ground state energy eigenvalues for the $x^4+iax$ potential with 
$a=5$ and 
$a=7$.  For these specific choices of $a$, the eigenfunction of the system does not respect the PT-symmetry as a result we are getting complex spectra. These plots show that the eigenvalues converge rapidly as the matrix size $K$ increases.



\section{Discussion} \label{sec7}
In quantum mechanical systems, the bootstrap technique has been applied to solve the eigenspectra of Hermitian Hamiltonians and extended to non-Hermitian PT-symmetric systems. However, progress in bootstrapping quantum systems has so far been confined to real spectra. Complex spectra naturally arise in open quantum systems, where the system interacts with its environment, but the standard bootstrap condition fails for non-Hermitian systems. In this work, we have identified suitable bootstrap conditions for a generic non-Hermitian system and used this technique to generate the eigenspectra of any complex polynomial potential, which includes PT-symmetric Hamiltonians as a special case. Furthermore, we have demonstrated that the method effectively addresses various boundary condition problems imposed on the eigenfunction, including the important application of capturing the PT-symmetric phase transition.

One of the key challenges in quantum systems is managing the increasing complexity and computational cost as the system size grows, particularly when calculating the spectra of the Lindbladian. These calculations become significantly more complex and time-consuming with larger systems. The  ``Bootstrap" technique we've developed has the potential to address this issue, especially in determining the Lindbladian gap of a dissipative system, which is crucial for understanding the thermalization time scale. Other important quantities, such as observables in the steady state, are typically difficult to compute. It would be interesting to apply the bootstrap technique to tackle these challenges. Additionally, our method is well-suited for identifying the PT-symmetric phase transition in spin-chain systems.
\begin{table}[h!]
    \centering
\begin{tabular}{ |p{2.7cm}||p{2.7cm}|p{2.7cm}|  }
 \hline
 \multicolumn{3}{|c|}{PT-symmetric phase transition} \\
 \hline
 $a$ &Ground state& K\\
 \hline
   1&   1.19 & 14\\
 & &  \\
 \hline
 3  & 2.62  & 16\\
 & &  \\
 \hline
  5  & 4.24+2.05$i$  & 17\\
 & &  \\
 \hline
  7  &  5.66+3.89$i$  & 15\\
 & &  \\
 \hline
\end{tabular}
\captionof{table}{This table indicates that for 
$a=1$ and 
$a=3$, the ground state eigenvalues are real. However, for 
$a=5$ and 
$a=7$, the ground state eigenvalues become complex because, for these values of $a$, the eigenfunction does not respect the PT-symmetry, demonstrating PT-symmetry breaking.}
\label{tab:t5}
\end{table}


\section*{Acknowledgements} 
SK acknowledges the CSIR fellowship with Grant Number 09/0936(11643)/2021-EMR-I. SK would like to thank M. Rangamani, R. Loganayagam, and others for the extensive discussions during the poster session at the FPQS conference held at IISER Pune. SK acknowledges M. Rangamani, D. Berenstein, and W. Li for reading the draft and providing useful suggestions. SK and HR would like to thank S. Jain for the valuable discussions.   
\bibliography{references.bib}
\bibliographystyle{apsrev4-2}


\appendix

\begin{widetext}

\section{ Bootstrapping conditions for non-Hermitian system}\label{app1}
    We take the Hilbert space $\mathcal{H}$ to be the space of square-integrable functions over the interval
   $x \in [-\infty, \infty]$ i.e. $L_{2}(\mathbb{R})$. Let's define the inner product on  $L_{2}(\mathbb{R})$ as
   \begin{equation}
       (\phi, \psi)=\int^{\infty}_{-\infty} dx \Bar{\phi}(x) \psi(x).
   \end{equation}
Note that, here $ \Bar{\phi}(x)$ denotes the complex conjugate of $ \Bar{\phi}(x)$. We take the Hamiltonian of the following form 
\begin{equation}\label{}
	H=p^2+V(x)=p^2+V_{1}(x)+ i\;
 V_{2}(x)\;.
\end{equation}
Since the Hamiltonian is not a self-adjoint operator, the left and the right eigenvector will be different. Let's assume, $R_{n}(x)$  is the n'th right eigenvector of the above Hamiltonian then we can write down the following eigenvalue equation
\begin{equation}
	HR_{n}(x)=(E^{n}_{\mathcal{R}} +iE^{n}_{\mathcal{I}}) R_{n}(x) .
\end{equation}
Using the
eigenvalue equation, we can write down  the following expression
\begin{align}\label{eqnhb}
	&    (H R_{n},OR_{n}) - (R_{n},OHR_{n})  =2 iE^{n}_{\mathcal{I}} (R_{n},OR_{n}) \nonumber\\
	&    (R_{n},OHR_{n})  =(E^{n}_{\mathcal{R}} +iE^{n}_{\mathcal{I}})(R_{n},OR_{n}) \;.
\end{align}
The above equation makes sense as long as the eigenvector, $ R_{n}(x)$, belongs to the Domain of operate $O$ i.e. $R_{n}(x) \in D(O)$.
By doing some algebraic manipulation, we can write the first part of the above equation in the following way
\begin{align}\label{eqnhb}
	&     (R_{n},(OH-H^{\dagger}O)R_{n}) +\mathcal{A} =2 iE^{n}_{\mathcal{I}} (R_{n},OR_{n})  \;,
\end{align}
where $H^\dagger$ is the complex conjugate of $H$ i.e. $H^\dagger=p^2+V_{1}(x)- i\;
 V_{2}(x)$ and the anomaly term is given by
 \begin{equation}
     \mathcal{A}=  (R_{n},H^\dagger O R_{n})- (H R_{n}, O R_{n})\;.
 \end{equation}
 In our case, we can explicitly show that the anomaly term vanishes for the choices $O=x^t$ and $O=x^tp$. Thus we recover the Eq.\eqref{bcn} presented in the main text.
 

  \begin{figure}
	\centering
 {\includegraphics[scale=.225]{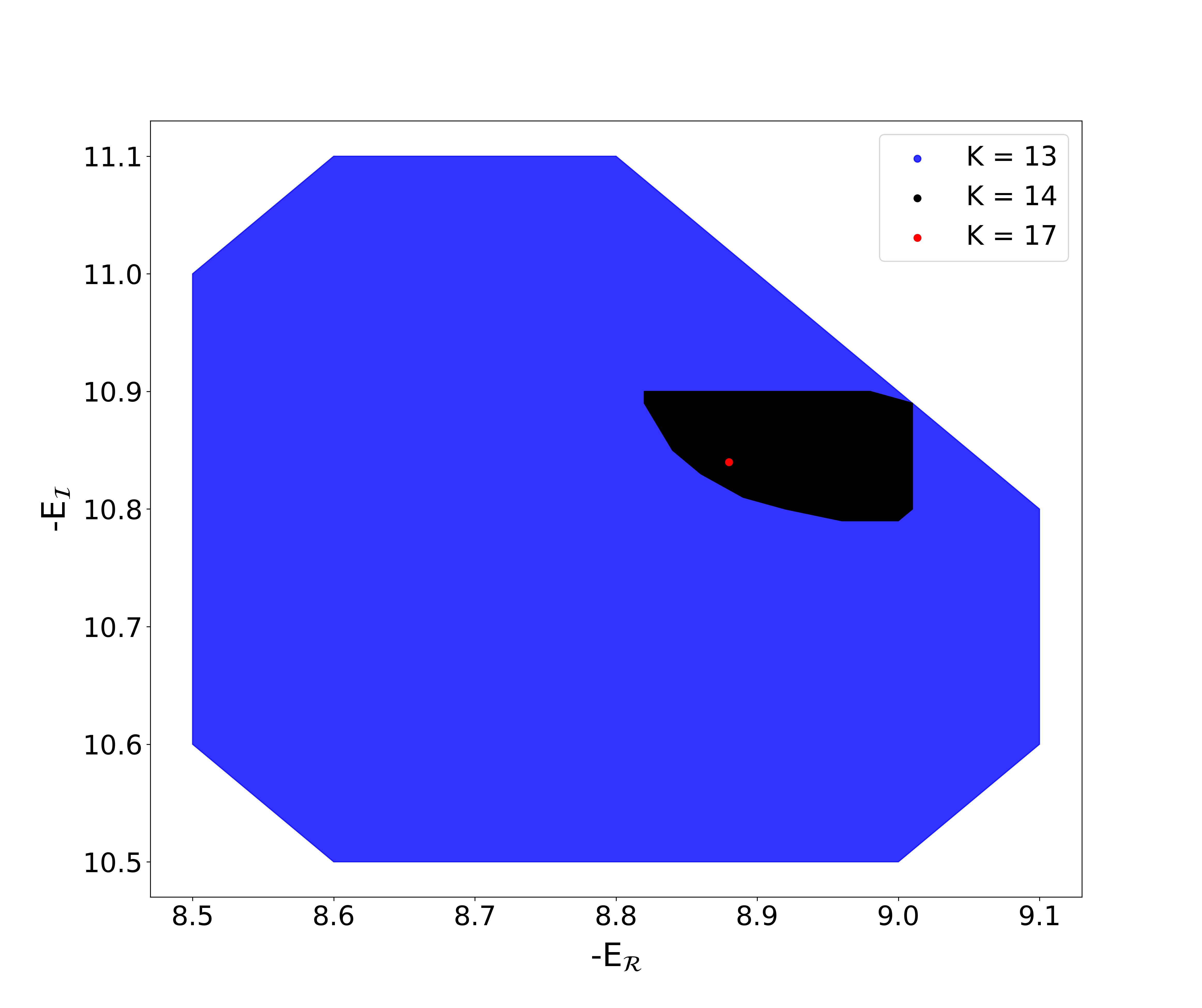}}
{\includegraphics[scale=.225]{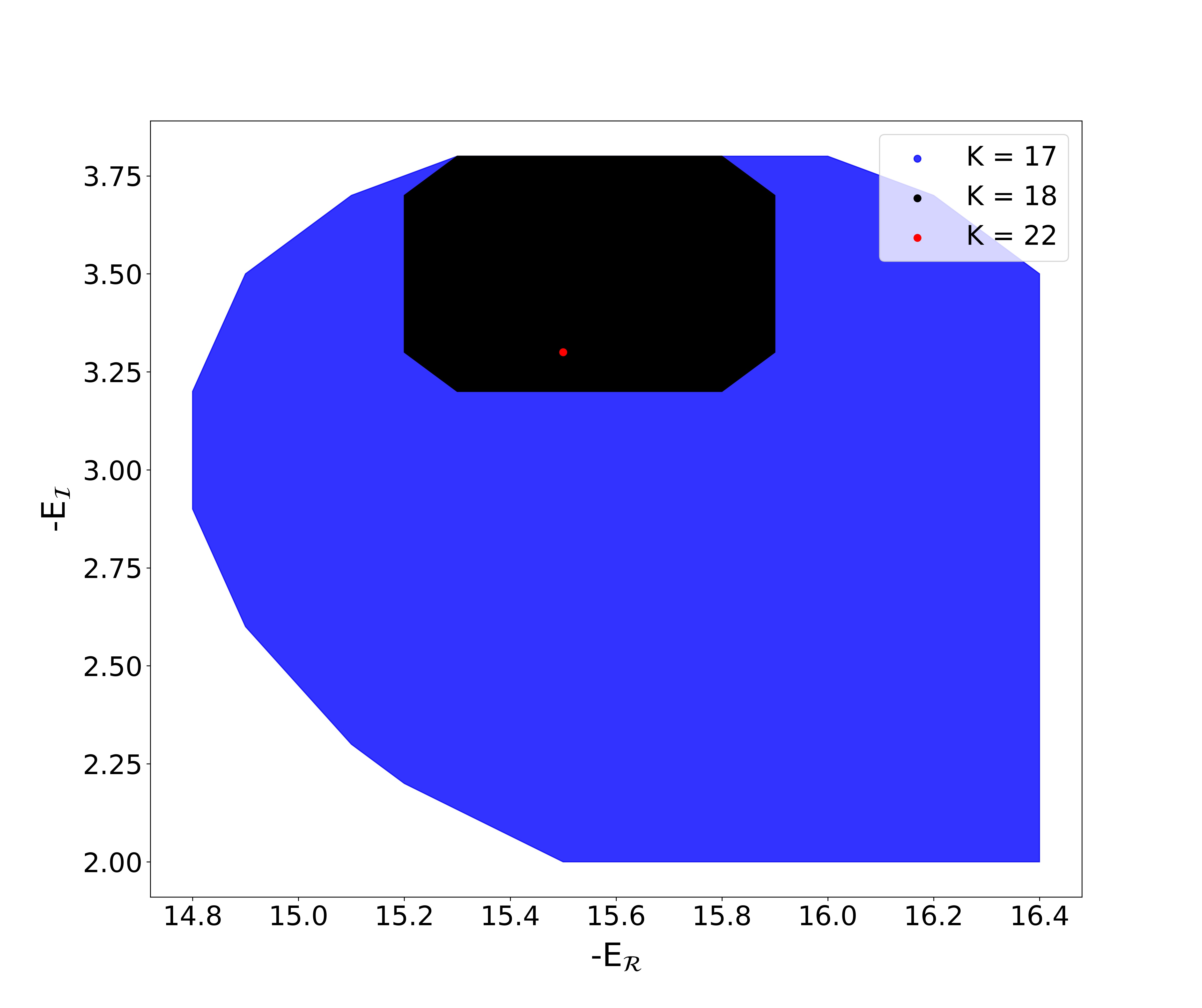}}
    \caption{
    (Color online:) We have plotted the first and second energy eigenvalues for the $x^6+15ix$ potential, utilizing different bootstrap matrix sizes $K$. Although the potential is PT-symmetric, the eigenvalues are complex because the eigenfunction does not respect the PT-symmetry. The plot shows that both the real and imaginary parts of the eigenvalues converge very fast as the matrix size $K$ increases. }
	\label{fig:4}
\end{figure}

     \begin{figure}
	\centering
 {\includegraphics[scale=.225]{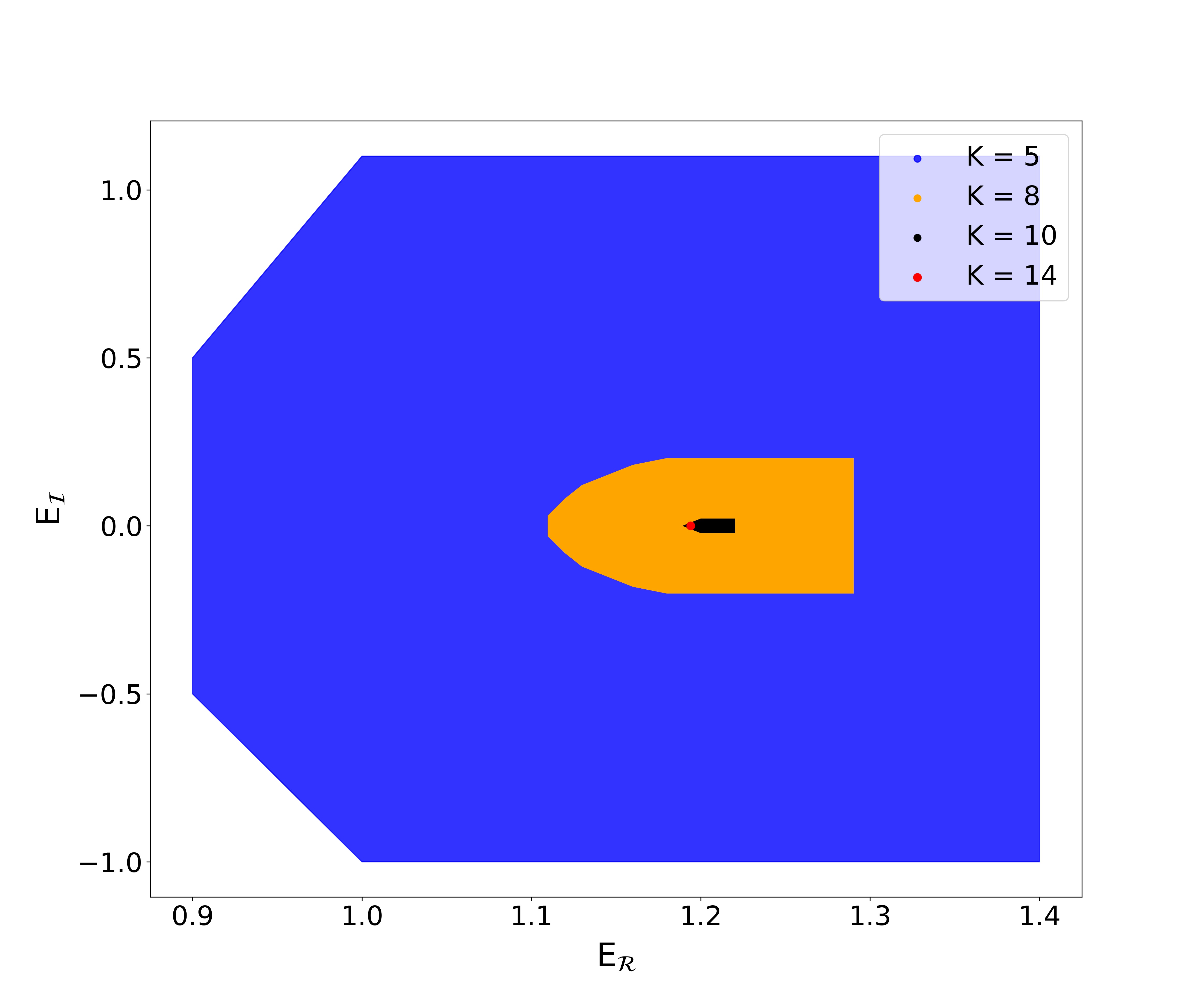}}
{\includegraphics[scale=.225]{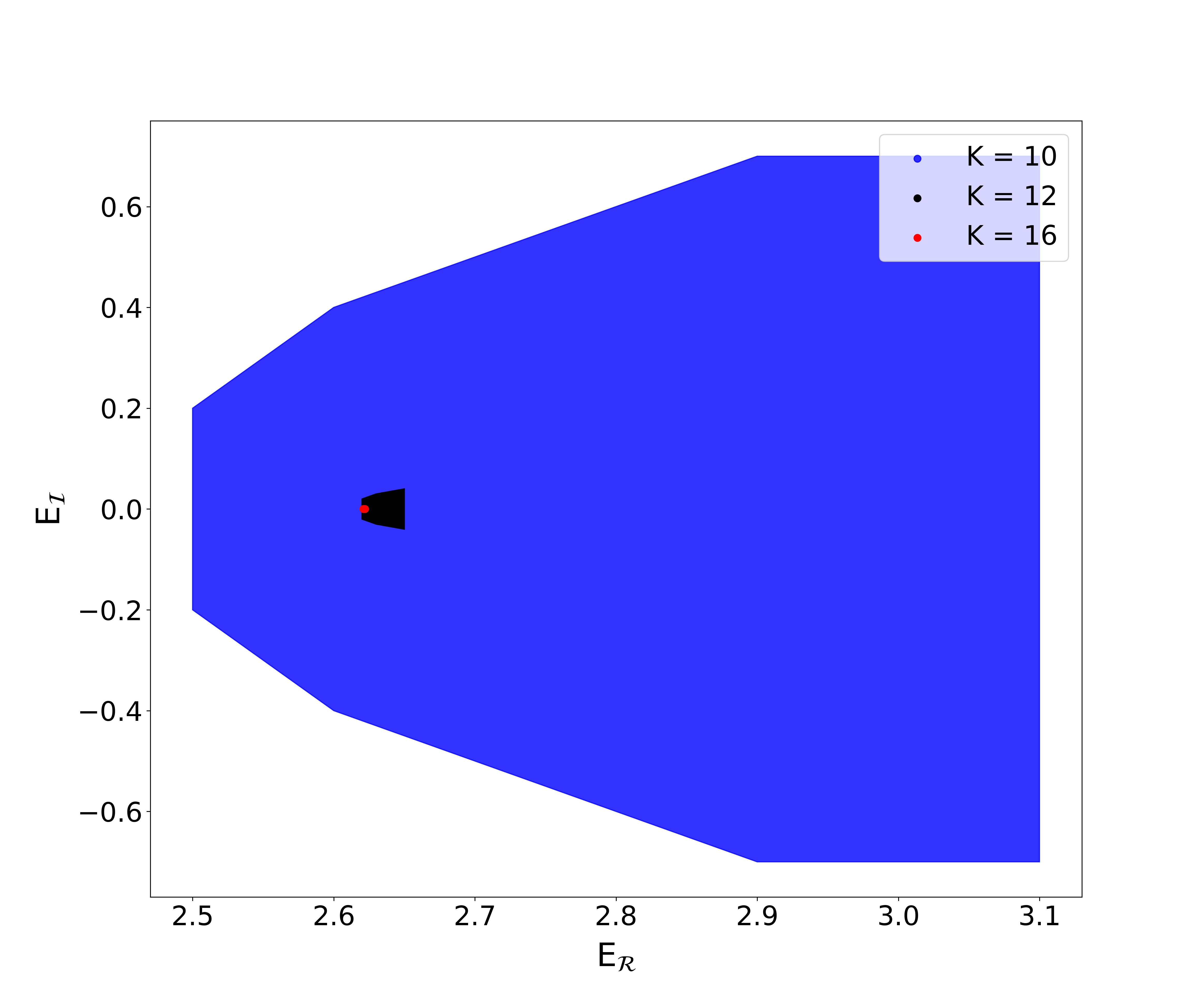}}
	\caption{(Color online:) We have plotted the ground state energy eigenvalues for the PT-symmetric $x^4+iax$ potential with 
$a=1$ and 
$a=3$, using various bootstrap matrix sizes 
$K$.  For $a=1$ and 
$a=3$, the eigenfunction of the system respects the PT-symmetry, ensuring the real spectra. The plot shows that both the eigenvalues converge very fast as the matrix size $K$ increases. }
	\label{fig:5}
\end{figure}

\begin{figure}
	\centering
{\includegraphics[scale=.225]{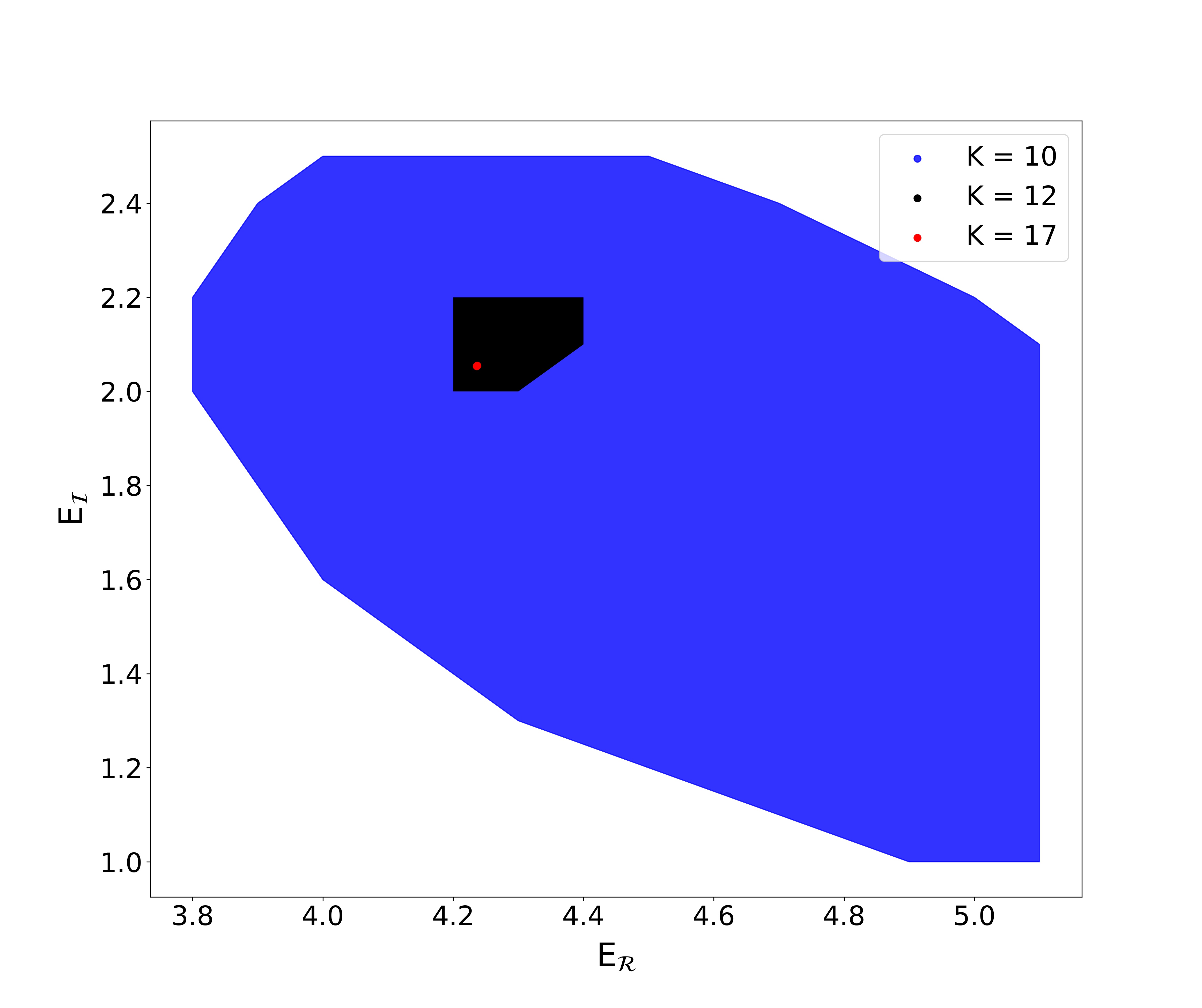}}
	{\includegraphics[scale=.054]{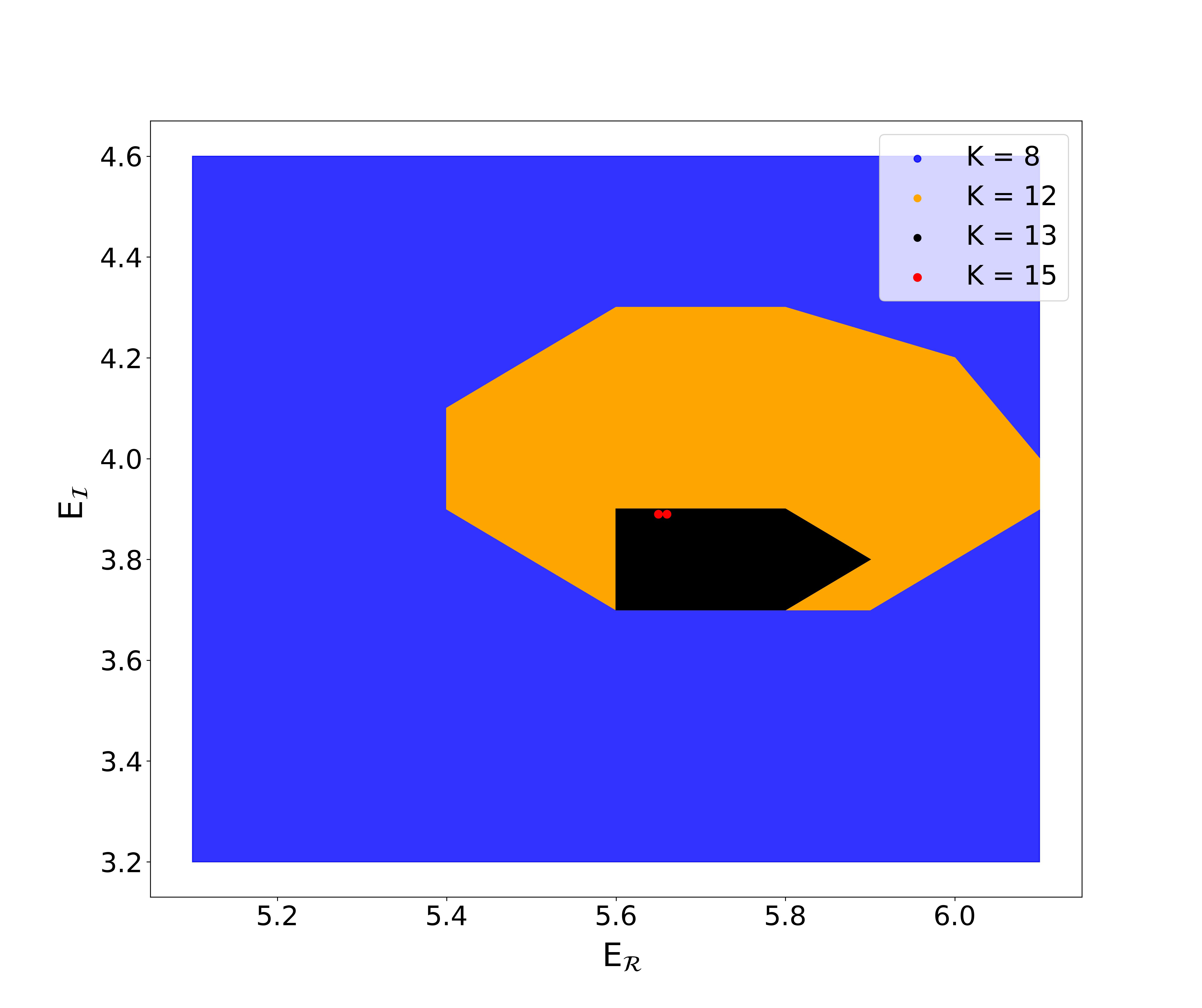}}
    \caption{
    (Color online:) We have plotted the ground state energy eigenvalues for the $x^4+iax$ potential with 
$a=5$ and 
$a=7$, using various bootstrap matrix sizes 
$K$.  For $a=5$ and 
$a=7$, the eigenfunction of the system does not respect the PT-symmetry as a result we are getting complex spectra. The plot shows that both the real and imaginary parts of the eigenvalues converge very fast as the matrix size $K$ increases.}
	\label{fig:6}
\end{figure}


\end{widetext}
 
\end{document}